# An Algorithm for Mining Multidimensional Fuzzy Assoiation Rules

[1]Neelu Khare
Department of computer Applications
MANIT, Bhopal    (M.P.)
Email: neelukh_29@yahoo.com

[2]Neeru Adlakha
Department of Applied Mathematics
SVNIT, Surat (Gujrat)
Email: neeru.adlakha21@gmail.com

[3]K. R. Pardasani
Department of Mathematics
MANIT, Bhopal (M.P.)
Email: kamalrajp@hotmail.com

**Abstract—** Multidimensional association rule mining searches for interesting relationship among the values from different dimensions/attributes in a relational database. In this method the correlation is among set of dimensions i.e., the items forming a rule come from different dimensions. Therefore each dimension should be partitioned at the fuzzy set level. This paper proposes a new algorithm for generating multidimensional association rules by utilizing fuzzy sets. A database consisting of fuzzy transactions, the Apriory property is employed to prune the useless candidates, itemsets.

*Keywords- interdimension ; multidimensional association rules; fuzzy membership functions ;categories.*

## I. INTRODUCTION

Data Mining is a recently emerging field, connecting the three worlds of Databases, Artificial Intelligence and Statistics. The computer age has enabled people to gather large volumes of data. Every large organization amasses data on its clients or members, and these databases tend to be enormous. The usefulness of this data is negligible if "meaningful information" or "knowledge" cannot be extracted from it. Data Mining answers this need.

Discovering association rules from large databases has been actively pursued since it was first presented in 1993, which is a data mining task that discovers associations among items in transaction databases such as the sales data [1]. Such kind of associations could be "if a set of items *A* occurs in a sale transaction, then another set of items *B* will likely also occur in the same transaction". One of the best studied models for data mining is that of *association rules* [2]. This model assumes that the basic object of our interest is an *item*, and that data appear in the form of sets of items called *transactions*. Association rules are "implications" that relate the presence of items in transactions [16]. The classical example is the rules extracted from the content of market baskets. Items are things we can buy in a market, and transactions are market baskets containing several items [17][18].

Association rules relate the presence of items in the same basket, for example, "every basket that contains bread contains butter", usually noted bread $\Rightarrow$ butter [3]. The basic format of an association rule is: An association is an implication of expression of the form A $\Rightarrow$ B, where A and B is disjoint itemset, i.e., A$\cap$B=$\phi$. The strength of an association rule can be measured in terms of its *support* and *confidence*. Support determines how often a rule is applicable to a given data set, while confidence determines how frequently items in B appear in transactions that contain A [5]. The formal definitions of these metrics are

Support   s (A $\Rightarrow$ B) = $\dfrac{\sigma(A \cup B)}{N}$

Confidence   c (A $\Rightarrow$ B) = $\dfrac{\sigma(A \cup B)}{\sigma(A)}$

In general, association rule mining can be viewed as a two-step process :

**1.** Find all frequent itemsets: By definition, each of these itemsets will occur at least as frequently as a predetermined minimum support count, min_sup.

**2.** Generate strong association rules from the frequent itemsets: By definition, these rules must satisfy minimum support and minimum confidence [6].

Association rule mining that implies a single predicate is referred as a single dimensional or intra-dimension association rule since it contains a single distinct predicate with multiple occurrences (the Predicate occurs more than once within the rule) [8]. The terminology of single dimensional or intra-dimension association rule is used in multidimensional database by assuming each distinct predicate in the rule as a dimension [11]. Association rules that involve two or more dimensions or predicates can be referred as multidimensional association rules. Rather than searching for frequent itemsets (as is done in mining single dimensional association rules), in multidimensional association rules, we search for frequent predicate sets (here the items forming a rule come from different dimensions) [10]. In general, there are two types of multidimensional association rules, namely inter-dimension association rules and hybrid-dimension association rules [15]. Inter-dimension association rules are multidimensional association rules with no repeated predicates. This paper introduces a method for generating inter-dimension association rules. Here, we introduce the concept of fuzzy transaction as a subset of items. In addition we present a general model to discover association rules in fuzzy transactions. We call them fuzzy association rules.

## II. APRIORI ALGORITHM AND APRIORI PROPERTY

Apriori is an influential algorithm in market basket analysis for mining frequent itemsets for Boolean association rules [1]. The name of Apriori is based on the fact that the algorithm uses prior knowledge of frequent itemset properties. Apriori employs an iterative approach known as a *level-wise* search, where *k*-itemsets are used to explore (*k+1*)-itemsets[2]. First, the set of frequent 1-itemsets is found, denoted by *L1*. *L1* is used to find *L2*, the set of frequent 2-itemsets, which is used to find *L3*, and so on, until no more frequent k-itemsets can be found.
Property: All non empty subsets of frequent item sets must be frequent [5].

### A. Itemsets in multidimensional data sets

Let RD is relational database with m records and n dimensions [4]. It consists of a set of all attributes/dimensions $D = \{d_1 \wedge d_2 \ldots \wedge d_n\}$ and set of tuples $T = \{t_1, t_2, t_3 \ldots t_m\}$ [10]. Where $t_i$ represents $i^{th}$ tuple and if there are n domains of attributes $D_1, D_2, D_3 \ldots D_n$, then each tuple $t_i = (v_{i1} \wedge v_{i2} \ldots \wedge v_{in})$ here $v_{ij}$ is atomic value of tuple $t_i$ with $v_{ij} \in D_j$ ; j-th value in i-th record, $1 \leq i \leq m$ and $1 \leq j \leq n$ [9]. Whereas RD can be defined as: $RD \subseteq D_1 \times D_2 \times D_3 \ldots \times D_n$.

To generate Multidimensional association rules we search for frequent predicate sets. A k-predicate set contains k conjunctive predicate. Dimensions, which are also called predications or fields, constitute a dimension combination with a formula (d1, d2, …, dn), in which dj represents j-th dimension [9]. The form $(d_j, v_{ij})$ is called an "item" in relational database or other multidimensional data sets, which is denoted by Iij. That is: $I_{ij} = (d_j, v_{ij})$, where $1 \leq i \leq m$ and $1 \leq j \leq n$. Suppose that A and B are items in the same relational database RD. A equals B if and only if the dimension and the value of the item A are equal to the dimension and the value of the item B, which is denoted by A=B. If it is not true that A equals B, then it is denoted by $A \neq B$. A set constituted by some "item" defined above is called "itemset".

## III. GENERATION OF FUZZY ASSOCIATION RULES

Apriori algorithm ignores the number items when determining relationship of the items. The algorithm that calculates support of itemsets just count the number of occurrences of the itemsets, in every record of transaction (shopping cart), without any consideration of the number of items in a record of transaction. However, based on *human intuitive*, it should be considered that the larger number of items purchased in a transaction means that the degree of association among the items in the transaction may be lower. When calculating support of itemsets in relational database, count of the number of categories/values in every dimension / attribute will be considered to find support of an item [8]. A new algorithm is proposed by considering that every item will have a relation (similarity) to the others if they are purchased together in the same record of transaction. They will have stronger relationship if they are purchased in more transactions. On the other hand, increasing number of categories/values in a dimension will reduce the total degree of relationship among the items involved from the different dimensions [12]. The proposed algorithm is given in the following steps:

**Step-1:**
Determine $\lambda \in \{2, 3, \ldots, n\}$(maximum value threshold). $\lambda$ is a threshold to determine maximum number of categories/values in a dimension by which the dimension can or cannot be considered in the process of generating rules mining. In this case, the process just considers all dimensions with the number of categories/values in the relational database RD less than or equal to $\lambda$. Formally, let $\mathbf{D_A} = (\mathbf{D_1} \times \mathbf{D_2} \times \mathbf{D_3} \ldots \times \mathbf{D_n})$ is a universal set of attributes or a domain of attributes/dimensions [13]. $\mathbf{M} \subseteq \mathbf{D_A}$ is a subset of qualified attributes/dimensions for generating rules mining that the number of unique categories/values(n) in $\mathbf{D_A}$ are not greater than $\lambda$:

$$\mathbf{M} = \{\mathbf{D} | n(\mathbf{D}) \leq \lambda \} \quad (1)$$

where n(D) is the number of categories/values in attribute/dimension D.

**Step-2:**
Set $k=1$, where $k$ is an index variable to determine the number of combination items in itemsets called *k*-itemsets. Whereas each item belongs to different attribute/dimension.

**Step-3:**
Determine minimum support for *k*-itemsets, denoted by $\beta_k \in (0, |M|)$ as a minimum threshold of a combination *k* items appearing from the whole qualified dimensions, where $|\mathbf{M}|$ is the number of qualified dimensions. Here, $\beta_k$ may have different value for every *k*.

**Step-4:**
Construct every candidate *k*-itemset, $I_k$, as a fuzzy set on set of qualified transactions, **M**. A fuzzy membership function, μ, is a mapping:

$\mu_{I^K} : M \rightarrow [0,1]$ as defined by:

$$\mu_{I^K}(D) = \inf_{i_{ij} \in D_j} \left\{ \frac{\eta_D(i_{ij})}{n(D_j)} \right\}, \forall D_j \in M \quad (2)$$

Where $I^K$ is a k-itemset, and items belongs to different dimensions.

A Boolean membership function, η, is a mapping:
$\eta_D : D \rightarrow \{0,1\}$ as defined by:

$$\eta_D(i) = \begin{cases} 1, i \in D \\ 0, otherwise \end{cases} \quad (3)$$

such that if an item, *i*, is an element of D then $\eta_D(i) = 1$, otherwise $\eta_D(i) = 0$.

**Step-5:**
Calculate *support* for every (candidate) *k*-itemset using the following equations [7]:

$$\text{Support}(I^K) = \sum_{T \in M} \mu_{I^K}(D) \quad (4)$$

**M** is the set of qualified dimensions as given in (1); it can be proved that (4) satisfied the following property:

$$\sum_{i \in D} \text{Support}(i) = |M|$$

For *k*=1, $I^k$ can be considered as a single item.

**Step-6:**
$I^k$ will be stored in the set of frequent *k*-itemsets, $L_k$ if and only if *support* $(I^k) \geq \beta_k$.

**Step-7:**
Set *k*=*k*+1, and if $k > \lambda$, then go to Step-9.

**Step-8:**
Looking for possible/candidate *k*-itemsets from $L_{k-1}$ by the following rules: A k-itemset, $I^k$, will be considered as a candidate *k*-itemset if $I^k$ satisfied:

$$\forall F \subset I^k, |F| = k-1 \Rightarrow F \in L$$

For example, $I_k = \{i_1, i_2, i_3, i_4\}$ will be considered as a candidate 4-itemset, iff: $\{i_1, i_2, i_3\}$, $\{i_2, i_3, i_4\}, \{i_1, i_3, i_4\}$ and $\{i_1, i_2, i_4\}$ are in $L_3$. If there is not found any candidate k-itemset then go to Step-9. Otherwise, the process is going to Step-3.

**Step-9:**
Similar to Apriori Algorithm, confidence of an association rule mining, $A \Rightarrow B$, can be calculated by the following equation[14]:

$$\text{Conf} = (A \Rightarrow B) = P(B|A) = \frac{\text{Support}(A \cup B)}{\text{Support}(A)} \quad (5)$$

where $A, B \in D_A$.

It can be followed that (5) can be also represented by:

$$\text{Conf}(A \Rightarrow B) = \frac{\sum_{D \in M} \inf_{i \in A \cup B}(\mu_i(D))}{\sum_{D \in M} \inf_{i \in A}(\mu_i(D))} \quad (6)$$

Where *A* and *B* are any *k*-itemsets in $L_k$. (Note: $\mu_i(T) = \mu_{\{i\}}(T)$, for simplification)[12]. Therefore, support of an itemset as given by (4) can be also expressed by:

$$\text{Support}(I^K) = \sum_{D \in M} \inf_{i \in I^K}(\mu_i(D)) \quad (7)$$

## IV. AN ILLUSTRATIVE EXAMPLE

An illustrative example is given to understand well the concept of the proposed algorithm and how the process of the generating fuzzy association rule mining is performed step by step. The process is started from a given relational database as shown in TABLE I.

TABLE I.

| TID | A  | B  | C  | D  | E  | F  |
|-----|----|----|----|----|----|----|
| T1  | A1 | B1 | C1 | D1 | E1 | F1 |
| T2  | A2 | B2 | C2 | D1 | E1 | F2 |
| T3  | A2 | B2 | C2 | D2 | E1 | F1 |
| T4  | A1 | B3 | C2 | D1 | E1 | F4 |
| T5  | A2 | B3 | C2 | D1 | E2 | F3 |
| T6  | A2 | B1 | C2 | D1 | E2 | F1 |
| T7  | A1 | B2 | C1 | D2 | E1 | F4 |
| T8  | A1 | B2 | C2 | D1 | E1 | F2 |
| T9  | A1 | B2 | C1 | D2 | E1 | F4 |
| T10 | A2 | B3 | C1 | D1 | E2 | F3 |

**Step-1:**
Suppose that $\lambda$ arbitrarily equals to 3; that means qualified attribute/dimension is regarded as an attribute/dimension with no more than 3 values/categories in the attribute/dimension. Result of this step is a set of qualified attribute/dimensions as seen in TABLE II.

TABLE II.

| TID | A  | B  | C  | D  | E  |
|-----|----|----|----|----|----|
| T1  | A1 | B1 | C1 | D1 | E1 |
| T2  | A2 | B2 | C2 | D1 | E1 |
| T3  | A2 | B2 | C2 | D2 | E1 |
| T4  | A1 | B3 | C2 | D1 | E1 |
| T5  | A2 | B3 | C2 | D1 | E2 |
| T6  | A2 | B1 | C2 | D1 | E2 |
| T7  | A1 | B2 | C1 | D2 | E1 |
| T8  | A1 | B2 | C2 | D1 | E1 |
| T9  | A1 | B2 | C1 | D2 | E1 |
| T10 | A2 | B3 | C1 | D1 | E2 |

where **M**={A,B,C,D,E}

**Step-2:**
The process is started by looking for support of 1-itemsets for which *k* is set equal to 1.

**Step-3:**
Since $\lambda$ =3 and $1 \leq k \leq m$. It is arbitrarily given $\beta_1$= 2, $\beta_2$ =2, $\beta_3$=1.5. That means the system just considers support of *k*-itemsets that is $\geq$ 2, for k=1,2 and $\geq$ 1.5, for k=3.

**Step-4:**
Every *k*-itemset is represented as a fuzzy set on set of transactions as given by the following results:
1-itemsets:
**{A1}** = {0.5/T1, 0.5/T4, 0.5/T7,0.5/T8, 0.5/T9},
**{A2}** = {0.5/T2, 0.5/T3, 0.5/T5, 0.5/T6, 0.5/T10},
**{B1}** = {0.33/T1, 0.33/T6},
**{B2}** = {0.33/T2, 0.33/T3, 0.33/T7, 0.5/T8, 0.5/T9,},
**{B3}** = {0.5/T4, 0.33/T5, 0.33/T10},
**{C1}** = {0.5/T1, 0.5/T7, 0.5/T9, 0.5/T10},
**{C2}** = {0.5/T2, 0.5/T3,0.5/T4, 0.5/T5, 0.5/T6, 0.5/T8},
**{D1}** = {0.5/T1, 0.5/T2, 0.5/T4,0.5/T5, 0.5/T6,0.5/T8, 0.5/T10},
**{D2}** = {0.5/T3, 0.5/T7, 0.5/T9},
**{E1}** = {0.5/T1, 0.5/T2, 0.5/T3, 0.5/T4, 0.5/T7,0.5/T8, 0.5/T9},
**{E2}** = {0.5/T5, 0.5/T6, 0.5/T10}

From Step-5 and Step-6, **{B1},{B2},{B3},{D2},{E2}** cannot be considered for further process their because *support* is < $\beta_1$.

2-itemsets:
{A1,C1}={0.5/T1, 0.5/T7,0.5/T9},
{A1,C2}={0.5/T4, 0.5/T8},
{A2,C1}={0.5/T10},
{A2,C2}={0.5/T2, 0.5/T3, 0.5/T5,0.5/T6},
{A1, D1}={0.5/T1, 0.5/T4, 0.5/T8},
{A2,D1}={ 0.5/T2, 0.5/T5, 0.5/T6, 0.5/T10},
{A1,E1}={ 0.5/T1, 0.5/T4, 0.5/T7, 0.5/T8, 0.5/T9},
{A2,E1}={ 0.5/T2, 0.5/T3},
{C1,D1}={ 0.5/T1,0.5/T$_{10}$},
{C2,D1}={ 0.5/T2, 0.5/T4, 0.5/T5, 0.5/T6, 0.5/T8},
{C1,E1}={ 0.5/T1, 0.5/T7, 0.5/T9},
{C2,E1}={ 0.5/T2, 0.5/T3, 0.5/T4, 0.5/T8},
{D1,E1}={ 0.5/T1, 0.5/T2, 0.5/T4,0.5/T8}

From Step-5 and Step-6 {A1,C1}, {A2,C1}, {A1,C2}, {A1,D1}, {A2,E1}, {C1,D1}, {C1,E1} cannot be considered for further process because their *support*< β$_2$.

| 1-itemsets | 2-itemsets |
|---|---|
| support({**A1**}) = 2.5, | support({**A1,C1**}) = 1.5 |
| support({**A2**}) = 2.5, | support({**A1,C2**}) = 0.5 |
| support({**B1**}) = 0.66, | support({**A2,C1**}) = 0.5 |
| support({**B2**}) = 1.65, | support({**A2,C2**}) = 2.5 |
| support({**B3**}) = 0.99, | support({**A1,D1**}) = 1.5 |
| support({**C1**}) = 2, | support({**A2,D1**}) = 2 |
| support({**C2**}) = 3, | support({**A1,E1**}) = 2.5 |
| support({**D1**}) = 3.5, | support({**A2,E1**}) = 1 |
| support({**D2**}) =1.5, | support({**C1,D1**}) = 1 |
| support({**E1**}) = 3.5 | support({**C2,D1**}) = 2.5 |
| support({**E2**}) = 1.5 | support({**C1,E1**}) = 1.5 |
|  | support({**C2,E1**}) = 2 |
|  | support({**D1,E1**}) = 2 |

3-itemsets:
{A2,C2,D1} = {0.5/T$_2$, 0.5/T$_5$,0.5/T$_6$},
{C2,D1,E1} = {0.5/T$_2$, 0.5/T$_4$,0.5/T$_8$}

**Step-5:**
Support of each k-itemset is calculated as given in the following results:

3-itemsets:
support({**A2,C1,E1**}) = 1.5
support({**C2,D1,E1**}) = 1.5

TABLE III : $L_1$(β$_1$=2)

| $L_1$ | |
|---|---|
| {*A1*} | 2.5 |
| {*A2*} | 2.5 |
| {*C1*} | 2 |
| {*C2*} | 2 |
| {*D1*} | 3.5 |
| {*E1*} | 3.5 |

TABLE IV : $L_2$ (β$_2$=2)

| $L_2$ | |
|---|---|
| {*A2*,C2} | 2.5 |
| {*A2*,D1} | 2 |
| {*A1*,E1} | 2.5 |
| {C2,D1} | 2.5 |
| {D1,E1} | 2 |
| {C2,*E1*} | 2 |

TABLE V : $L_3$ (β$_3$=1.5)

| $L_3$ | |
|---|---|
| {*A2*,C2,D1} | 1.5 |
| {C2,D1,E1} | 1.5 |

**Step-6:**
From the results as performed by Step-4 and 5, the sets of frequent 1-itemsets, 2-itemsets and 3-itemsets are given in Table 8, 9 and 10, respectively.

**Step-7:**
This step is just for increment the value of k in which if all elements of $L_K$< β$_K$, then the process is going to Step-9.

**Step-8:**
This step is looking for possible/candidate *k*-itemsets from $L_{k-1}$. If there is no anymore candidate k-itemset then go to Step-9. Otherwise, the process is going to Step-3.

**Step-9:**
The step is to calculate every confidence of each possible association rules as follows:

$$Conf(A2 \Rightarrow C2) = \frac{Support(A2,C2)}{Support(A2)} = \frac{2.5}{2.5} = 1$$

.
.
.

$$Conf(A2 \Rightarrow D1) = \frac{Support(A2,D1)}{Support(A2)} = \frac{2}{2.5} = 0.8$$

$$Conf(A2 \Rightarrow C2 \wedge D1) = \frac{Support(A2,C2,D1)}{Support(A2)} = \frac{1.5}{2.5} = 0.6$$

$$Conf(C2 \wedge D1 \Rightarrow E1) = \frac{Support(C2,D1,E1)}{Support(C2,D1)} = \frac{1.5}{2.5} = 0.6$$

$$Conf(C2 \Rightarrow D1 \wedge E1) = \frac{Support(C2,D1,E1)}{Support(C2)} = \frac{1.5}{3} = 0.5$$

V. CONCLUSION

This paper introduced an algorithm for generating fuzzy multidimensional association rules mining as a generalization of inter-dimension association rule. The algorithm is based on the concept that the larger number of values/categories in a dimension/attribute means the lower degree of association among the items in the transaction. Moreover, to generalize inter-dimension association rules, the concept of fuzzy itemsets is discussed, in order to introduce the concept of fuzzy multidimensional association rules. Two generalized formulas were also proposed in the relation to the fuzzy association rules. Finally, an illustrated example is given to clearly demonstrate and understand steps of the algorithm.

In future we discuss and propose a method to generate conditional hybrid dimension association rules using fuzzy logic, whereas hybrid dimension association rule is hybridization between inter-dimension and intra-dimension association rules.

REFERENCES

[1] Agrawal, R., Imielinski, T., and Swami, A. N. "Mining association rules between sets of items in large databases". In Proceedings of the ACM SIGMOD International Conference on Management of Data, pp .207-216 (1993).

[2] Agrawal, R. and Srikant, R. "Fast algorithms for mining association rules". In Proc. 20th Int. Conf. Very Large Data Bases, pp. 487-499, (1994).

[3] Klemetinen, L., Mannila, H., Ronkainen, P., "Finding interesting rules from large sets of discovered association rules". Third International Conference on Information and Knowledge Management Gaithersburg, pp.401-407 USA (1994).

[4] Houtsma M, Swami A. "Set-oriented mining of association rules in relational databases". In: Proc of the 11th International Conference on Data Engineering. Taipei, pp. 25-33 Taiwan: 1995.

[5] R. Agrawal, A. Arning, T. Bollinger, M. Mehta,J. Shafer, and R. Srikant. " The Quest Data Mining System" , Proceedings of the 2nd Int'l Conference on Knowledge Discovery in Databases and Data Mining", Portland, Oregon, August (1996).

[6] J. Han, M. Kamber, Data Mining: Concepts and Techniques, The Morgan Kaufmann Series, (2001).

[7] G. J. Klir, B. Yuan, Fuzzy Sets and Fuzzy Logic: Theory and Applications, New Jersey: Prentice Hall, (1995).

[8] Jurgen M. Jams Fakultat ,"An Enhanced Apriori Algorithm for Mining Multidimensional Association Rules", 25th Int. Conf. Information Technology interfaces ITI Cavtat, Croatia (1994).

[9] Wan-Xin Xu1, Ru-Jing Wang, "A Fast Algorithm Of Mining Multidimensional Association Rules Frequently", Proceedings of the Fifth International Conference on Machine Learning and Cybernetics, Dalian, 13-16 August 2006 IEEE.

[10] Rolly Intan, Department of Informatics Engineering, Petra Christian University, Surabaya, "A Proposal Of Fuzzy Multidimensional Association Rules", Jurnal INFORMATIKA VOL 7: pp. 85-90 ,NOV 2006.

[11] Reda ALHAJJ ADSA , Mehmet KAYA ,"Integrating Fuzziness into OLAP for Multidimensional Fuzzy Association Rules Mining", Third IEEE International Conferenceon Data Mining (ICDM'03) , (2003)

[12]Rolly Intan, "An Algorithm for Generating Single Dimensional Fuzzy Association Rule Mining", *JURNAL INFORMATIKA VOL. 7, NO. 1, MEI 2006: 61 - 66* (2006)

[13] Rolly Intan, "Mining Multidimensional Fuzzy Association Rules from a Normalized Database", International Conference on Convergence and Hybrid Information Technology © 2008 IEEE

[14] Rolly Intan1, Oviliani Yenty Yuliana, Andreas Handojo, Mining "Multidimensional Fuzzy Association Rules From A Database Of Medical Record Patients" , Jurnal Informatika Vol. 9, No. 1, 15 - 22 Mei 2008.

[15] Anjna Pandey and K. R. Pardasani, "Rough Set Model for Discovering Multidimensional Association Rules"  IJCSNS VOL 9, pp 159-164,  June 2009.

[16] Miguel Delgado, Nicolás Marín, Daniel Sánchez, and María-Amparo Vila, "Fuzzy Association Rules: General Model and Applications" IEEE TRANSACTIONS ON FUZZY SYSTEMS, VOL. 11, NO. 2, APRIL 2003.

[17] Han, J., Pei, J., Yin, Y. "Mining Frequent Patterns without Candidate Generation",SIGMOD Conference, pp 1-12, ACM Press (2000).

[18] Han, J., Pei, J., Yin, Y., Mao, R. "Mining Frequent Patterns without Candidate Generation: A Frequent-Pattern Tree Approach". Data Mining and Knowledge Discovery,  53–87 (2004).

[19] Hannes Verlinde, Martine De Cock, and Raymond Boute, "Fuzzy Versus Quantitative Association Rules: A Fair Data-Driven Comparison"  IEEE TRANSACTIONS ON SYSTEMS, MAN, AND CYBERNETICS—PART B: CYBERNETICS, VOL. 36, NO. 3, pp.679-684, JUNE 2006.